\documentclass[runningheads]{lncse}
\usepackage{multicol} 
\usepackage{makeidx}  
\usepackage{amsmath}
\usepackage{graphicx}
\begin{document}
\input epic.sty
\input eepic.sty
\title{Evaluation of Single Vehicle Data in Depen- dence of
the Vehicle-Type, Lane, and Site}
\titlerunning{Evaluation of Single Vehicle Data}
\author{Benno Tilch\inst{1} \and Dirk Helbing\inst{1,2}}
\institute{II. Institute of Theoretical Physics, 
University of Stuttgart, Pfaffenwaldring 57/III, 70550 Stuttgart, Germany
\and
Collegium Budapest~-- Institute for Advanced Study,
H-1014 Hungary}
\maketitle
\abstract{\protect In this paper we study dependencies of fundamental diagrams, 
time gap distributions, and velocity-distance relations on vehicle
types, lanes and/or
measurement sites. We also propose measurement and aggregation methods that 
have more favourable statistical properties than conventional 
methods.\\[-1.0cm] \mbox{ }}

\section*{ }

In the recent two years, traffic flow modelling has been more and more
stimulated by empirical studies. In particular, single-vehicle data
may shed some more light on the mechanisms of the transition from
free to congested traffic flow. 
In contrast to early studies of,
for example, the distribution of time gaps between successive cars 
(for an overview see \cite{May90}), Bovy and coworkers \cite{Bovy98} have
carried out separate analyses for different sample periods
(morning/noon/evening) and different vehicle types (passenger-cars,
articulated and non-articulated trucks). In addition, they have
determined the relations between vehicle speeds and the distance
gaps to the respective car in front separately for free and congested
traffic, and for different cross sections of a Dutch freeway. Neubert, 
Santen, Schadschneider and Schreckenberg \cite{NeuSanSchaSchr99} 
mainly restrict to the analysis of one cross section of a 
German freeway. However, they go one step beyond the Dutch study
by distinguishing not only free and congested traffic, but also
different density regimes, which leads them to interesting conclusions.
In the following, we try to combine both
efforts by distinguishing free and congested traffic, different cross
sections, and different vehicle types as well.
\par
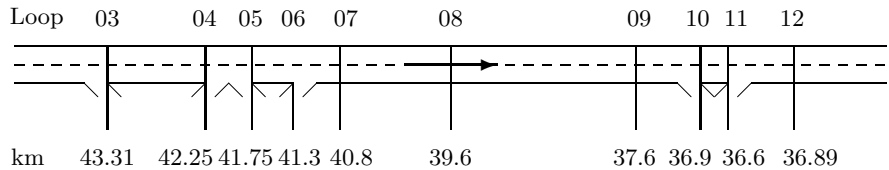
\begin{figure}[b]
  \begin{center}
    \unitlength1.23cm
\begin{picture}(9.42,1.3)(0,0.3)
  \thicklines
  \put(4.21,1.2){\vector(1,0){1.0}}
  \thinlines
  \put(0,1.4){\line(1,0){9.42}}

  \put(0,1.2){\line(1,0){0.1}}
  \put(0.2,1.2){\line(1,0){0.1}}
  \put(0.4,1.2){\line(1,0){0.1}}
  \put(0.6,1.2){\line(1,0){0.1}}
  \put(0.8,1.2){\line(1,0){0.1}}
  \put(1.0,1.2){\line(1,0){0.1}}
  \put(1.2,1.2){\line(1,0){0.1}}
  \put(1.4,1.2){\line(1,0){0.1}}
  \put(1.6,1.2){\line(1,0){0.1}}
  \put(1.8,1.2){\line(1,0){0.1}}
  \put(2.0,1.2){\line(1,0){0.1}}
  \put(2.2,1.2){\line(1,0){0.1}}
  \put(2.4,1.2){\line(1,0){0.1}}
  \put(2.6,1.2){\line(1,0){0.1}}
  \put(2.8,1.2){\line(1,0){0.1}}
  \put(3.0,1.2){\line(1,0){0.1}}
  \put(3.2,1.2){\line(1,0){0.1}}
  \put(3.4,1.2){\line(1,0){0.1}}
  \put(3.6,1.2){\line(1,0){0.1}}
  \put(3.8,1.2){\line(1,0){0.1}}
  \put(4.0,1.2){\line(1,0){0.1}}
  \put(5.3,1.2){\line(1,0){0.1}}
  \put(5.5,1.2){\line(1,0){0.1}}
  \put(5.7,1.2){\line(1,0){0.1}}
  \put(5.9,1.2){\line(1,0){0.1}}
  \put(6.1,1.2){\line(1,0){0.1}}
  \put(6.3,1.2){\line(1,0){0.1}}
  \put(6.5,1.2){\line(1,0){0.1}}
  \put(6.7,1.2){\line(1,0){0.1}}
  \put(6.9,1.2){\line(1,0){0.1}}
  \put(7.1,1.2){\line(1,0){0.1}}
  \put(7.3,1.2){\line(1,0){0.1}}
  \put(7.5,1.2){\line(1,0){0.1}}
  \put(7.7,1.2){\line(1,0){0.1}}
  \put(7.9,1.2){\line(1,0){0.1}}
  \put(8.1,1.2){\line(1,0){0.1}}
  \put(8.3,1.2){\line(1,0){0.1}}
  \put(8.5,1.2){\line(1,0){0.1}}
  \put(8.7,1.2){\line(1,0){0.1}}
  \put(8.9,1.2){\line(1,0){0.1}}
  \put(9.1,1.2){\line(1,0){0.1}}
  \put(9.3,1.2){\line(1,0){0.1}}

  \put(1,0.5){\line(0,1){0.9}}
  \put(2.06,0.5){\line(0,1){0.9}}
  \put(2.56,0.5){\line(0,1){0.9}}
  \put(3.01,0.5){\line(0,1){0.5}}
  \put(3.51,0.5){\line(0,1){0.9}}
  \put(4.71,0.5){\line(0,1){0.9}}
  \put(6.71,0.5){\line(0,1){0.9}}
  \put(7.41,0.5){\line(0,1){0.9}}
  \put(7.71,0.5){\line(0,1){0.9}}
  \put(8.42,0.5){\line(0,1){0.9}}
  \put(0.14,0.21){\makebox(0,0){\small km}}
  \put(1.00,0.2){\makebox(0,0){\small 43.31}}
  \put(1.85,0.2){\makebox(0,0){\small 42.25}}
  \put(2.50,0.2){\makebox(0,0){\small 41.75}}
  \put(3.08,0.2){\makebox(0,0){\small 41.3}}
  \put(3.64,0.2){\makebox(0,0){\small 40.8}}
  \put(4.71,0.2){\makebox(0,0){\small 39.6}}
  \put(6.70,0.2){\makebox(0,0){\small 37.6}}
  \put(7.30,0.2){\makebox(0,0){\small 36.9}}
  \put(7.88,0.2){\makebox(0,0){\small 36.6}}
  \put(8.60,0.2){\makebox(0,0){\small 36.89}}
 
  \put(0.24,1.7){\makebox(0,0){\small Loop}}
  \put(1.00,1.7){\makebox(0,0){\small 03}}
  \put(2.05,1.7){\makebox(0,0){\small 04}}
  \put(2.55,1.7){\makebox(0,0){\small 05}}
  \put(3.01,1.7){\makebox(0,0){\small 06}}
  \put(3.58,1.7){\makebox(0,0){\small 07}}
  \put(4.71,1.7){\makebox(0,0){\small 08}}
  \put(6.75,1.7){\makebox(0,0){\small 09}}
  \put(7.38,1.7){\makebox(0,0){\small 10}}
  \put(7.80,1.7){\makebox(0,0){\small 11}}
  \put(8.40,1.7){\makebox(0,0){\small 12}}
  \put(0,1){\line(1,0){0.75}}
  \put(0.75,1){\line(1,-1){0.15}}
  \put(1,1){\line(1,-1){0.15}}
  \put(1,1){\line(1,0){1.06}}
  \put(2.06,1){\line(-1,-1){0.15}}
  \put(2.31,1){\line(-1,-1){0.15}}
  \put(2.31,1){\line(1,-1){0.15}}
  \put(2.56,1){\line(1,-1){0.15}}
  \put(2.56,1){\line(1,0){0.45}}
  \put(3.01,1){\line(-1,-1){0.15}}
  \put(3.26,1){\line(-1,-1){0.15}}
  \put(3.26,1){\line(1,0){3.9}}
  \put(7.16,1){\line(1,-1){0.15}}
  \put(7.41,1){\line(1,-1){0.15}}
  \put(7.41,1){\line(1,0){0.3}}
  \put(7.71,1){\line(-1,-1){0.15}}
  \put(7.96,1){\line(-1,-1){0.15}}
  \put(7.96,1){\line(1,0){1.46}}
\end{picture}

    \caption{Investigated section of the Dutch freeway A9 from Haarlem 
to Amsterdam, where a speed limit of 120\,km/h applies,
and locations of the measurement sites.\label{fig:loops}}
  \end{center}
\end{figure}
Our single-vehicle data are from the Dutch highway A9 from Haarlem to
Amsterdam for the two periods 
10/10--10/14 and 10/31--11/04/1994, but we have neglected the time
intervals between midnight and 3 am. 
At several cross section of this freeway (see Fig.~\ref{fig:loops}),
double induction loops record the time of passing, 
the lane, the velocity, and the length of each passing vehicle.
In the following, vehicles longer than 6\,m are denoted as trucks,
shorter vehicles as cars. On average, there were about 20\% of trucks
in the right lane, and less than 1\% trucks in the left lane, but the 
proportion was strongly varying in time \cite{TreHel99}.
\par
\begin{figure}[b]
\begin{center}
     \includegraphics[scale=0.42]{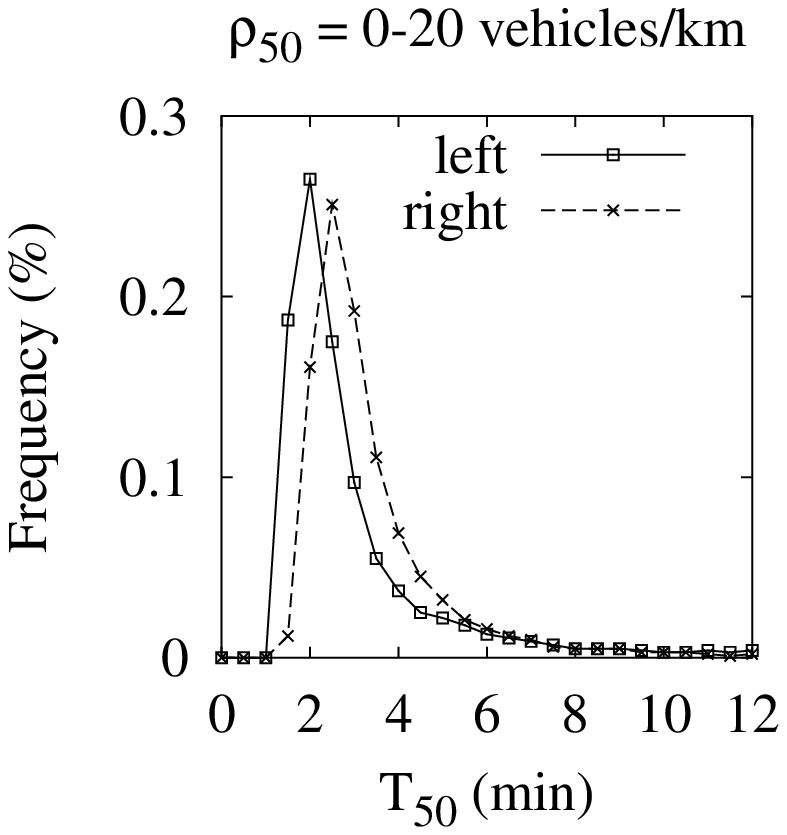}
     \includegraphics[scale=0.42]{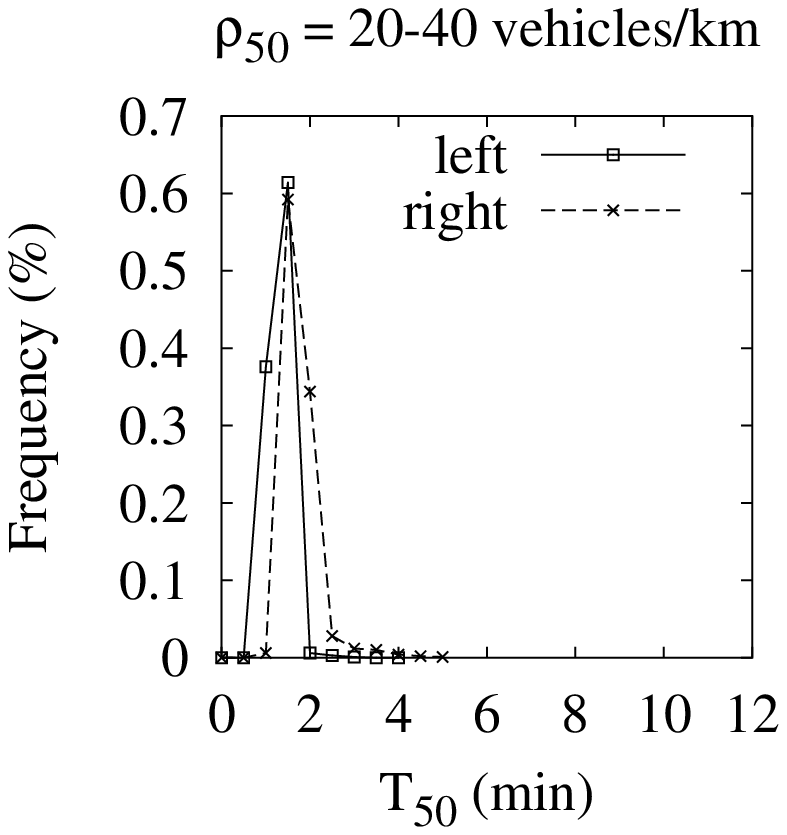}
     \includegraphics[scale=0.42]{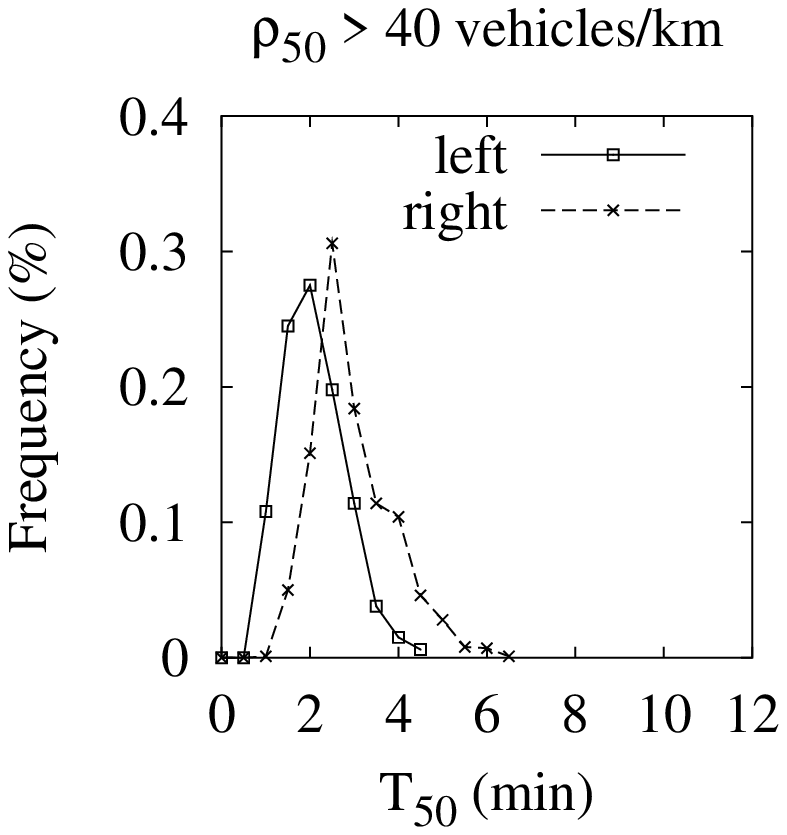}
     \caption{Distribution of the measurement intervals $T_{50}$ 
at different vehicle densities $\rho_{50}$. (The data were taken from
all loops.)\label{fig:T_50}}
\end{center}
\end{figure}

Although this is not significant for the results of this study,
we have not, as usual, 
determined macroscopic quantities like the traffic flow, 
the average velocity, or the vehicle density by averaging over
fixed time periods (like 1 minute). In order to have
comparable sample sizes, we have instead averaged over a fixed number
$N$ of cars, as suggested in Ref.~\cite{Hel97}.
Otherwise the {\em statistical} error at small traffic flows (i.e. at small and
large densities) would be quite large. This is compensated for by 
a flexible measurement interval $T_N$ (see Fig.~\ref{fig:T_50}). 
It is favourable 
that $T_N$ becomes particularly small in the (medium) density range of unstable
traffic, so that the method yields a good representation of traffic
dynamics. However, choosing small values of $N$ does not make sense,
since the temporal variation of the aggregate values will mainly reflect
statistical variations, then.
In order to have a time resolution of about 2 minutes on each lane,
one should select $N=50$, while $N=100$ can be chosen when averaging
over both lanes. Aggregate values over both lanes for $N=50$ are
comparable with 1-minute averages, but show a smaller statistical
scattering at low densities (compare results in \cite{Hel97}
and \cite{HelBook}). With increasing $N$, the maxima move to higher
values, and the distributions of $T_N$ become broader. 
The distribution for the left lane has its maxima at lower values of
$T_N$, probably because of the smaller number of trucks.
Throughout this paper, we use $N=50$, but we have checked that our
results don't change significantly for $N=30$ or $N=100$. 
\par
\begin{figure}[t]
  \begin{center}
     \includegraphics[scale=0.42]{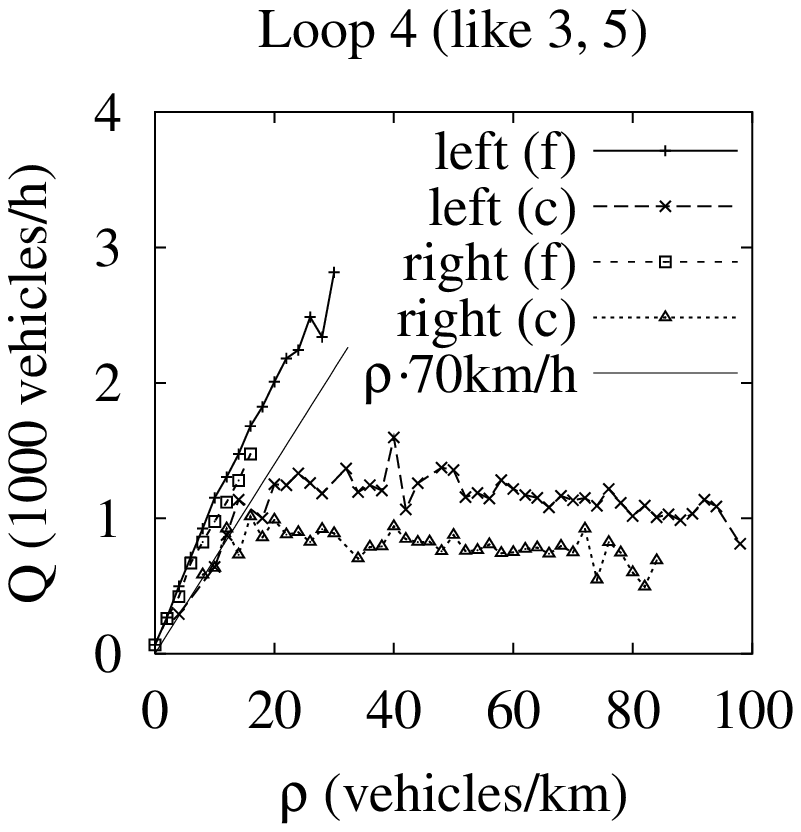}
     \includegraphics[scale=0.42]{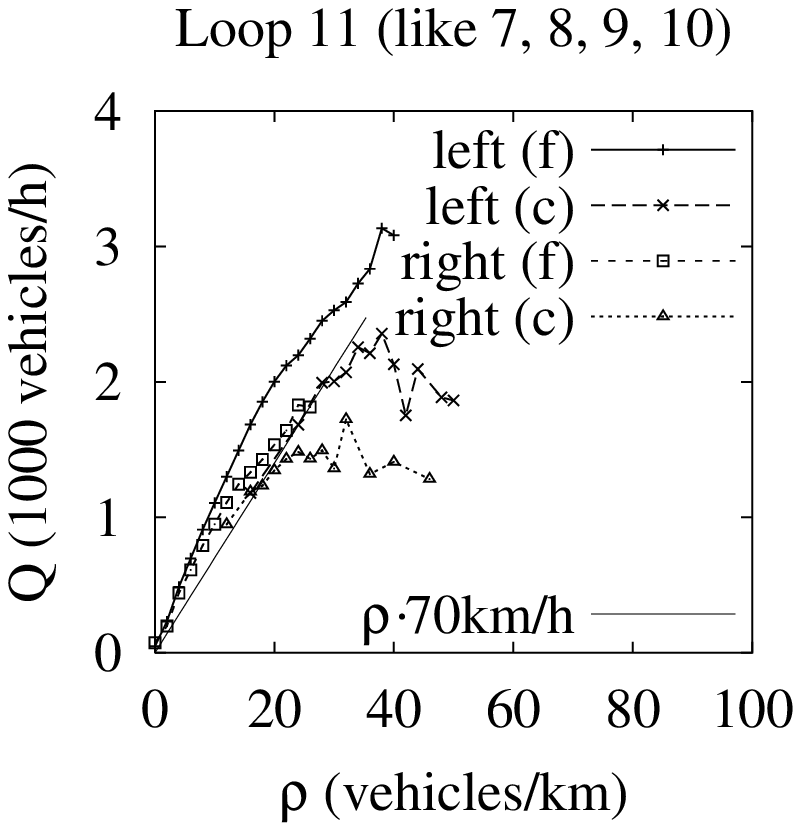}
     \includegraphics[scale=0.42]{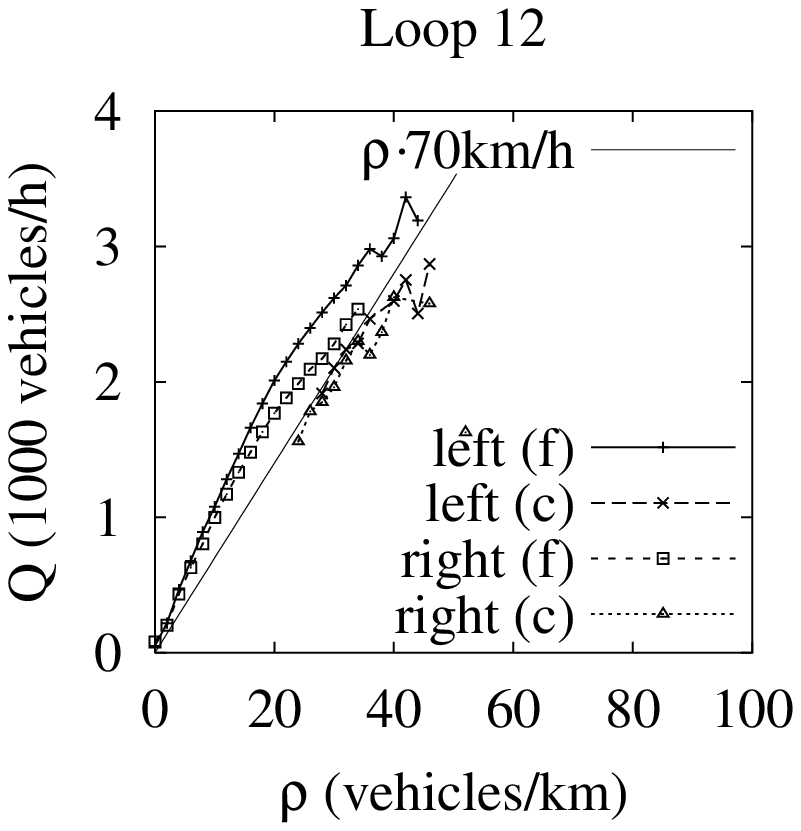}
     \caption[]{\protect Free (f) and congested (c) traffic flow as a
function of the density for the left and the right lane at different 
cross sections of the road: upstream (left), downstream (middle),
and far enough away from an on-ramp or bottleneck
(right). From left to right we observe a relaxation from congested to
free traffic.\\[-12mm]\mbox{ }\label{fig:flows}}
\end{center}
\end{figure}

Based on the passing times $t_i$ of successive vehicles $i$ in the same 
lane, we are able to calculate the time gaps $\Delta t_i = (t_i -
t_{i-1}) > 0$. 
The (measurement) time interval\\[-4mm]
\begin{equation}
  \label{eq:T_N}
  T_N = \sum_{i=i_0+1}^{i_0+N} (t_i - t_{i-1}) \equiv
\sum_{i=i_0+1}^{i_0+N} \Delta t_i
\end{equation}
for the passing of $N$ vehicles defines the (inverse of the)
traffic flow $Q_N$ by:\\[-4mm]
\begin{equation}
  \label{eq:Q_N}
  \frac{1}{Q_N} = \frac{T_N}{N} = \frac{1}{N} \sum_{i=i_0+1}^{i_0+N} 
  \Delta t_i \, .
\end{equation}
Note that the traffic flow is very much dependent on the measurement site
(Fig.~\ref{fig:flows}). Consequently, the following graphs are
different for other sites as well, but only in the quantitative
details, not in a qualitative (fundamental) way.
\par
\begin{figure}
  \begin{center}
     \includegraphics[scale=0.42]{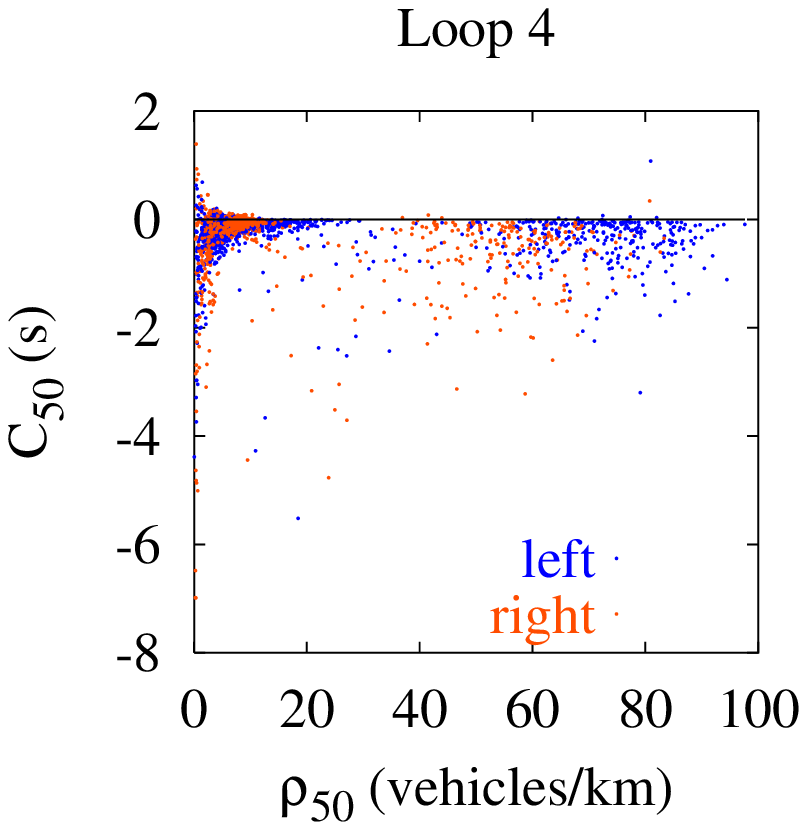}
     \includegraphics[scale=0.42]{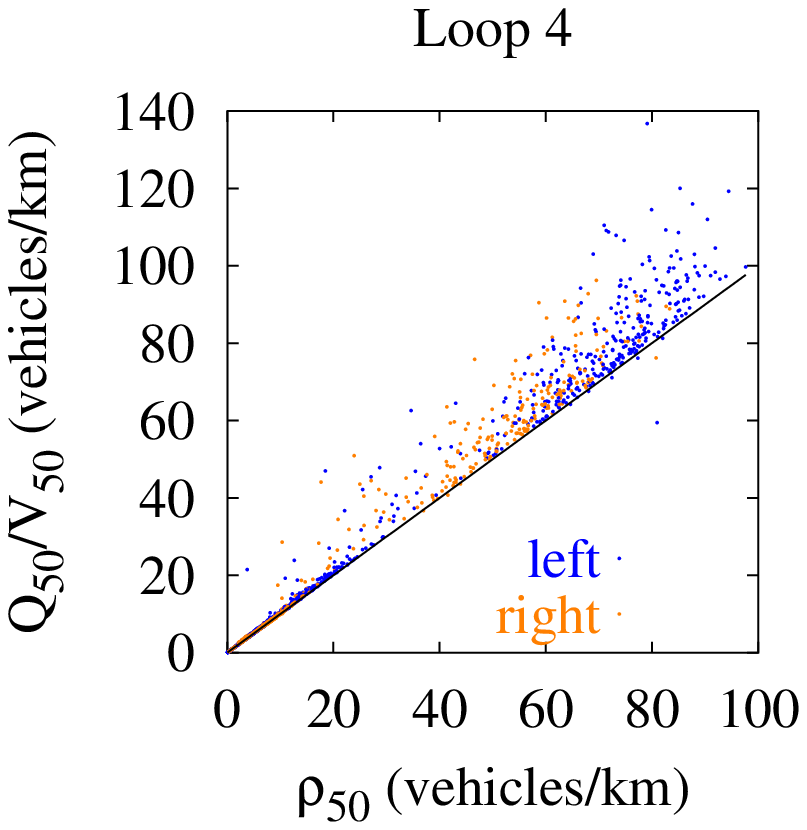}
     \includegraphics[scale=0.42]{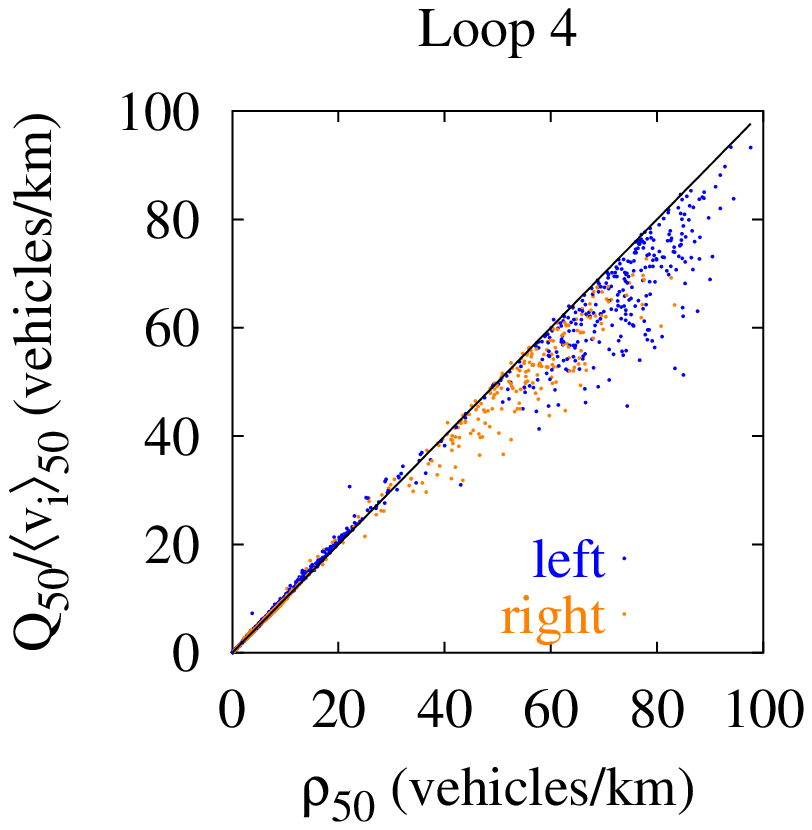}
     \caption[]{\protect 
Correlation $C_{50}$, density $Q_{50}/V_{50}$ according to the
fluid-dynamic formula, and conventionally determined density
$Q_{50}/\langle v_i\rangle_{50}$ as a function of the 
density $\rho_{50}$ according to the proposed definition (\ref{zzz}),
in comparison with the usually assumed relations 
(---).\\[-12mm] \mbox{ }\label{fig:C_rho}}
  \end{center}
\end{figure}

Now, we approximate the (brutto) distance gap $\Delta x_i$ 
by $\Delta x_i = v_i \Delta t_i$, 
where $v_i$ is the actual velocity of vehicle $i$.
This assumes that the vehicle velocities do not considerably change
during the time interval $\Delta t_i$ and implies
\begin{equation}
  \label{eq:cont}
  \frac{1}{Q_N} = \left\langle \Delta t_i \right\rangle_N 
  = \left\langle \frac{\Delta x_i}{v_i} \right\rangle_{\!N} 
  = \left\langle\Delta x_i \right\rangle_{N} \left\langle \frac{1}{v_i} 
  \right\rangle_{\!N} + C_N \, ,
\end{equation}
where $C_N$ is the covariance between the distance gaps $\Delta x_i$
and the inverse velocities $1/v_i$. We expect that this covariance 
is particularly relevant at large vehicle densities
(Fig.~\ref{fig:C_rho}, left). The
density $\rho_N$ and the average velocity $V_N$ are defined by
\begin{equation}
  \frac{1}{\rho_N^{ }} = \left\langle\Delta x_i \right\rangle_{N}
  = \frac{1}{N} \sum_{i=i_0+1}^{i_0+N} \Delta x_i
  \qquad\mbox{ and }\qquad
  \frac{1}{V_N} = \left\langle \frac{1}{v_i} \right\rangle_{\!N}
  = \frac{1}{N}  \sum_{i=i_0+1}^{i_0+N} \frac{1}{v_i} \, .
\label{zzz}
\end{equation}
Then, we obtain the fluid-dynamic flow relation $Q_N = \rho_N V_N$ by the
conventional
assumption $C_N = 0$ which, however, tends to
overestimate the density (Fig.~\ref{fig:C_rho}, middle). 
The fact that $V_N$ is defined as the harmonic mean value
of the vehicle velocities $v_i$ automatically corrects for the fact
that the {\em spatial} velocity distribution differs from the {\em
locally measured} one (see \cite{Leu88,HelBook} for details). The common
method of determining the density via $Q_N/\langle v_i\rangle_N$
underestimates the density (Fig.~\ref{fig:C_rho}, right).
%
\par
\begin{figure}
  \begin{center}
\includegraphics[scale=0.42]{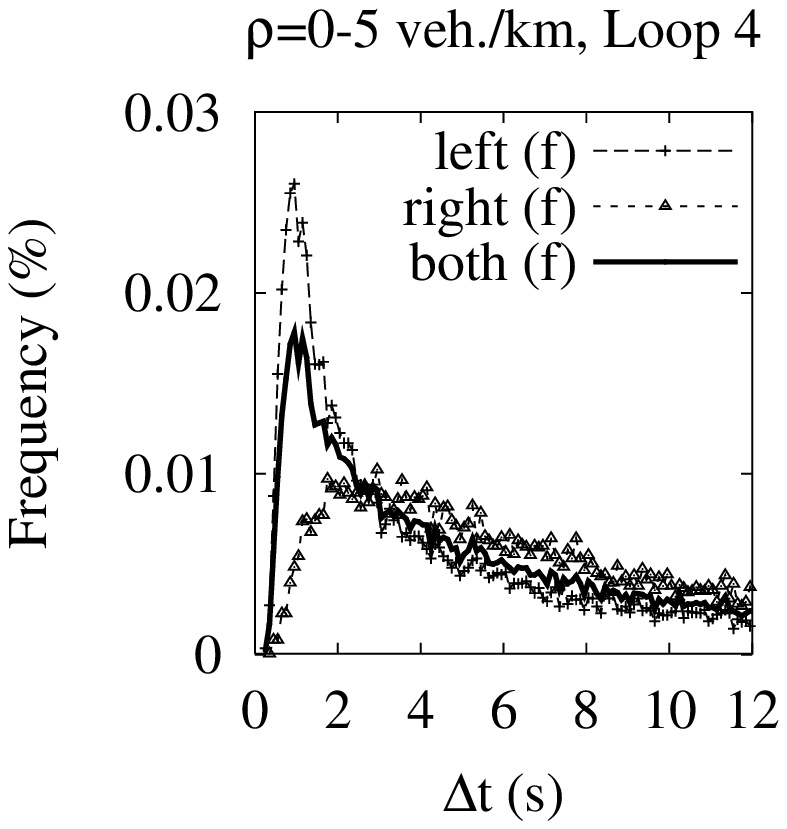}
\includegraphics[scale=0.42]{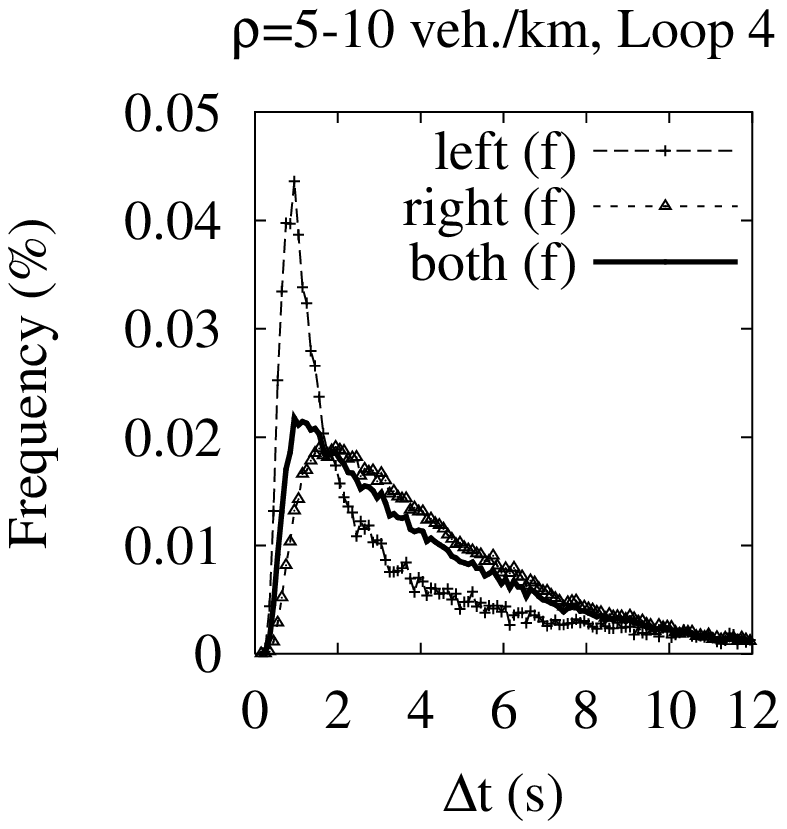}
\includegraphics[scale=0.42]{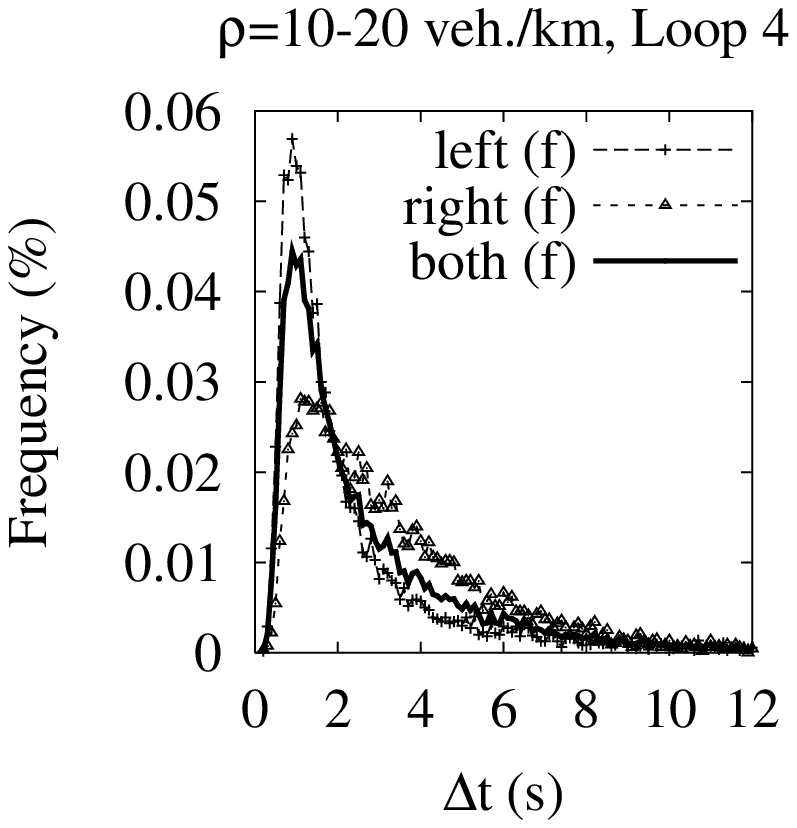}\\
\includegraphics[scale=0.42]{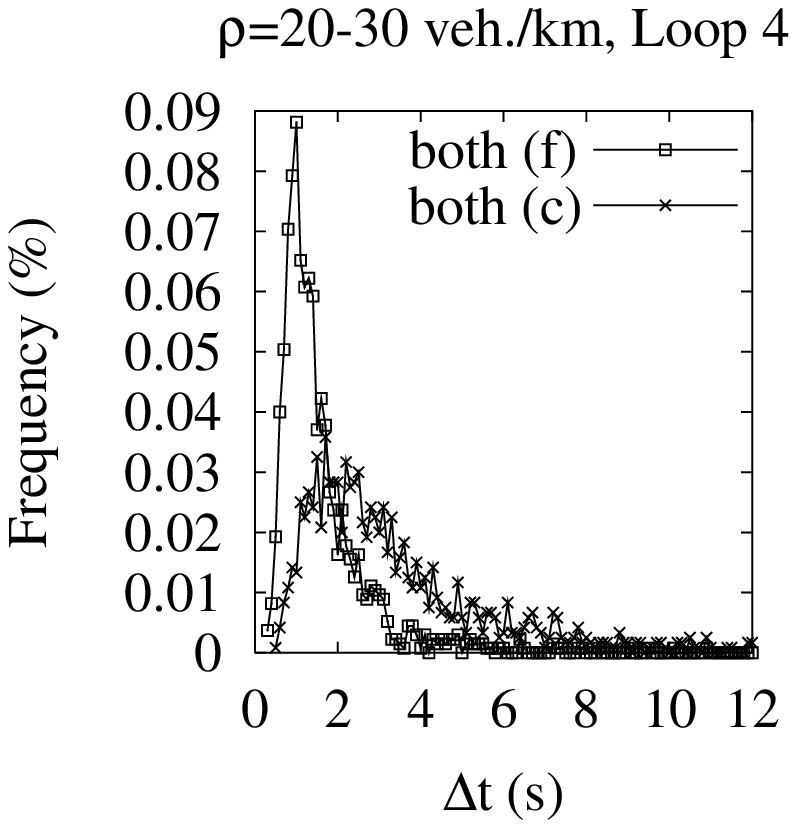}
\includegraphics[scale=0.42]{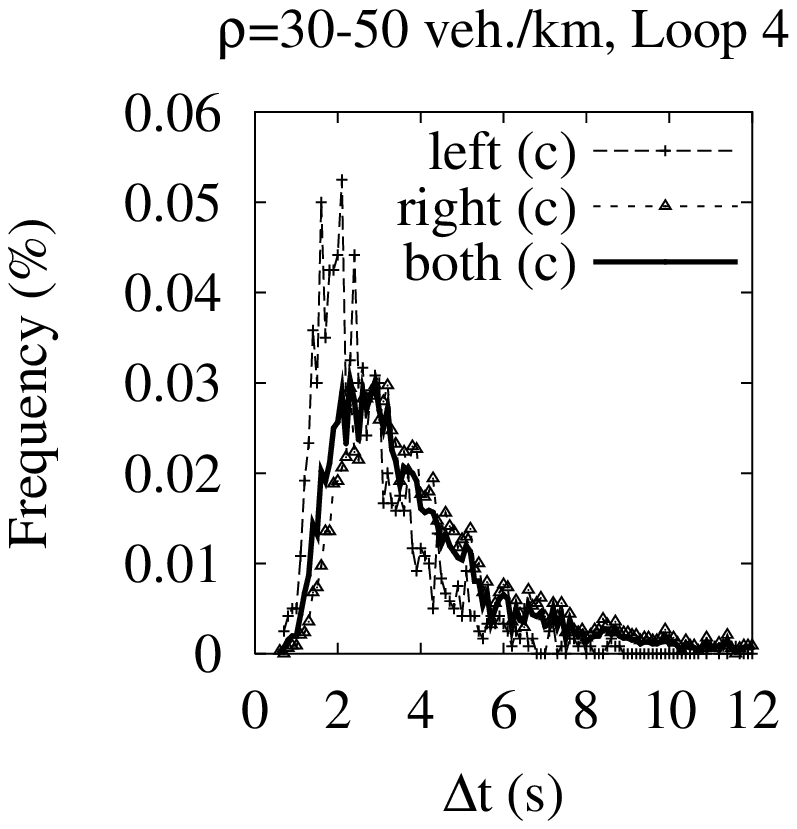}
\includegraphics[scale=0.42]{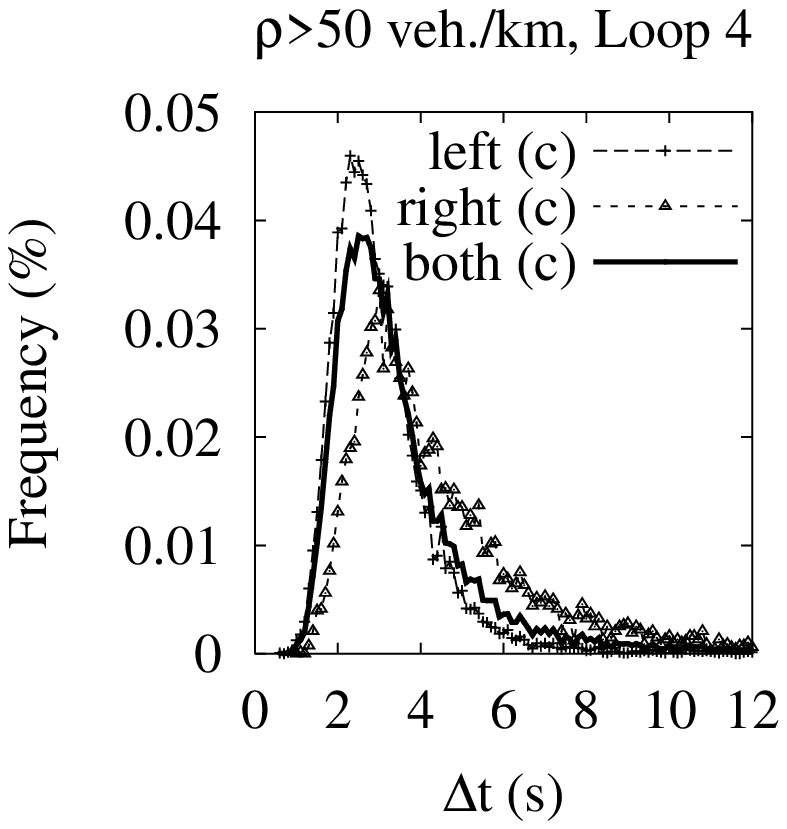}\\
\includegraphics[scale=0.42]{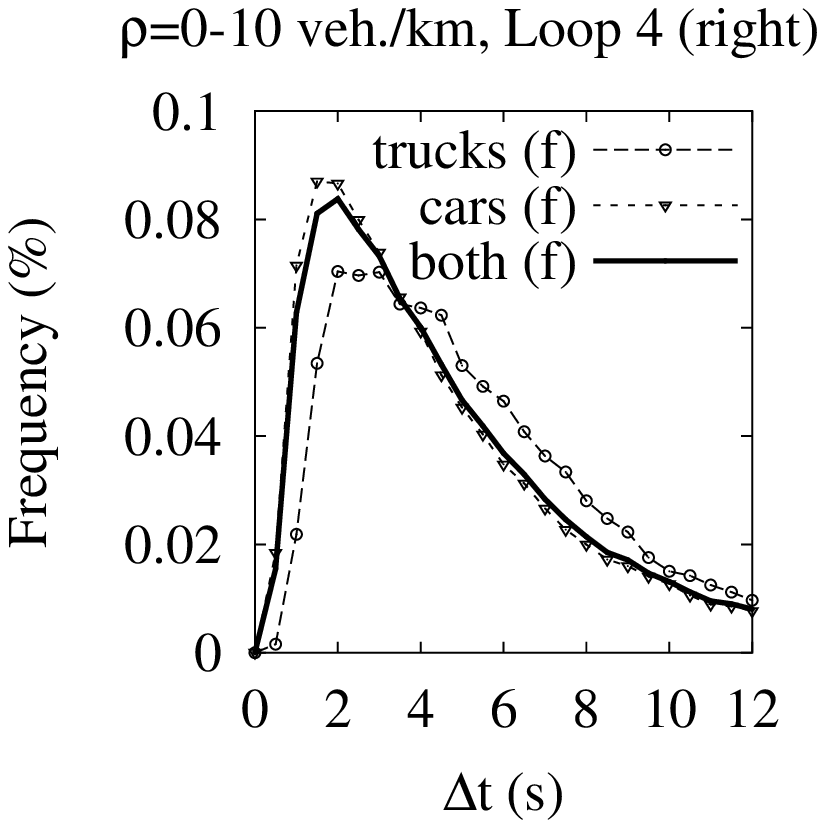}
\includegraphics[scale=0.42]{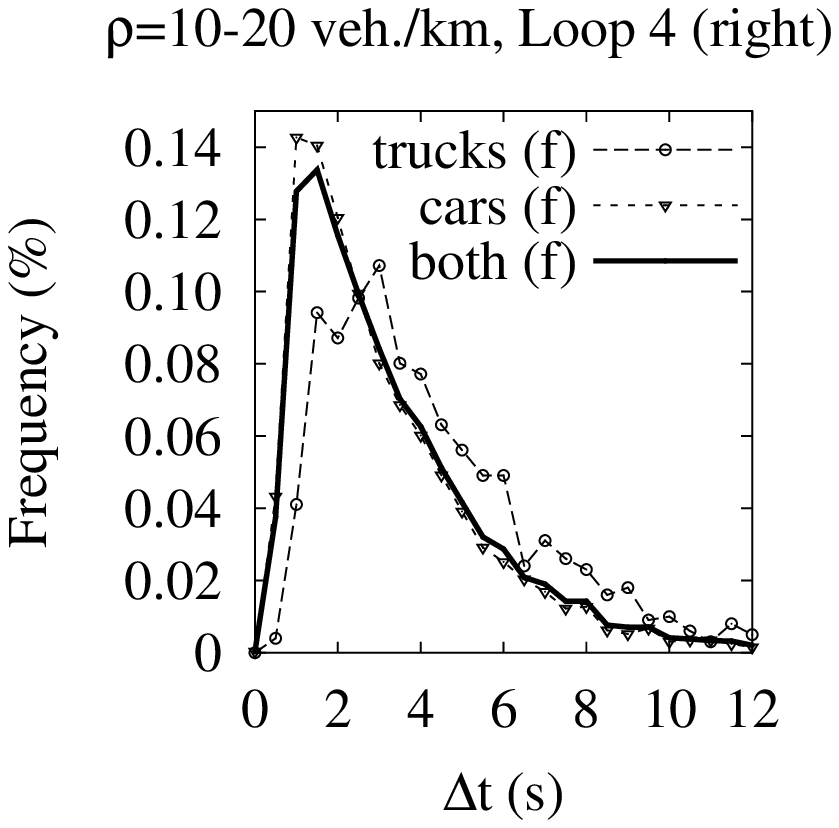}
\includegraphics[scale=0.42]{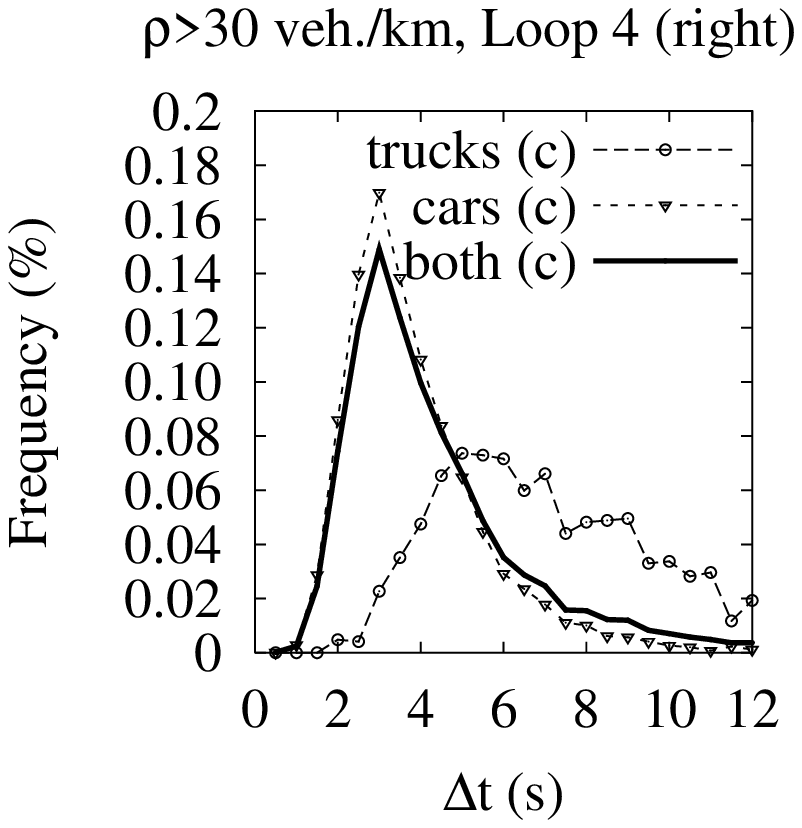}
    \caption{Time gap distributions in
different density regimes for different lanes and vehicle
types.}
\label{fig:tg}
\end{center}
\end{figure}
In our investigation of {\em time gap distributions}, we have not only
distinguished different density regimes, but also free traffic (f)
and congested traffic (c) (Fig.~\ref{fig:tg}). Measurement intervals with
$V_N < 70$\,km/h were classified as congested, otherwise traffic was
considered to be free.
\par
\begin{figure}[t]
  \begin{center}
    \includegraphics[scale=0.455]{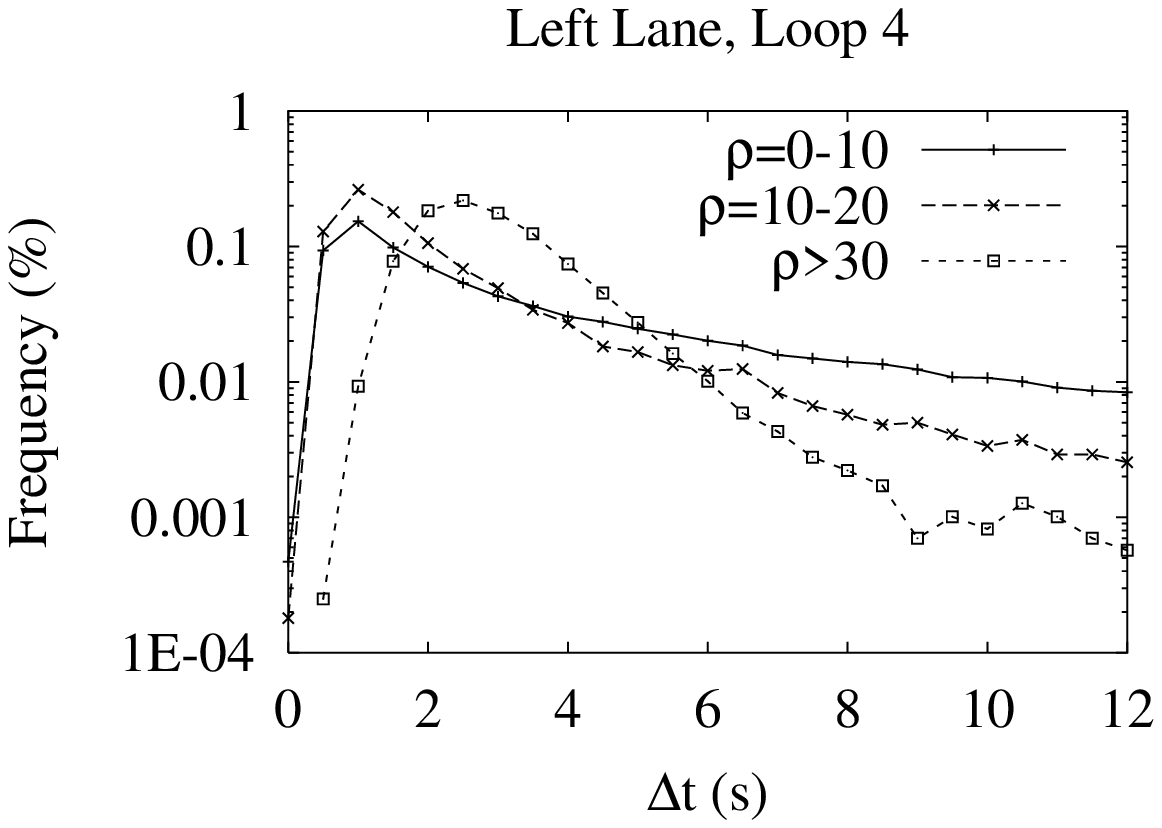}
    \includegraphics[scale=0.455]{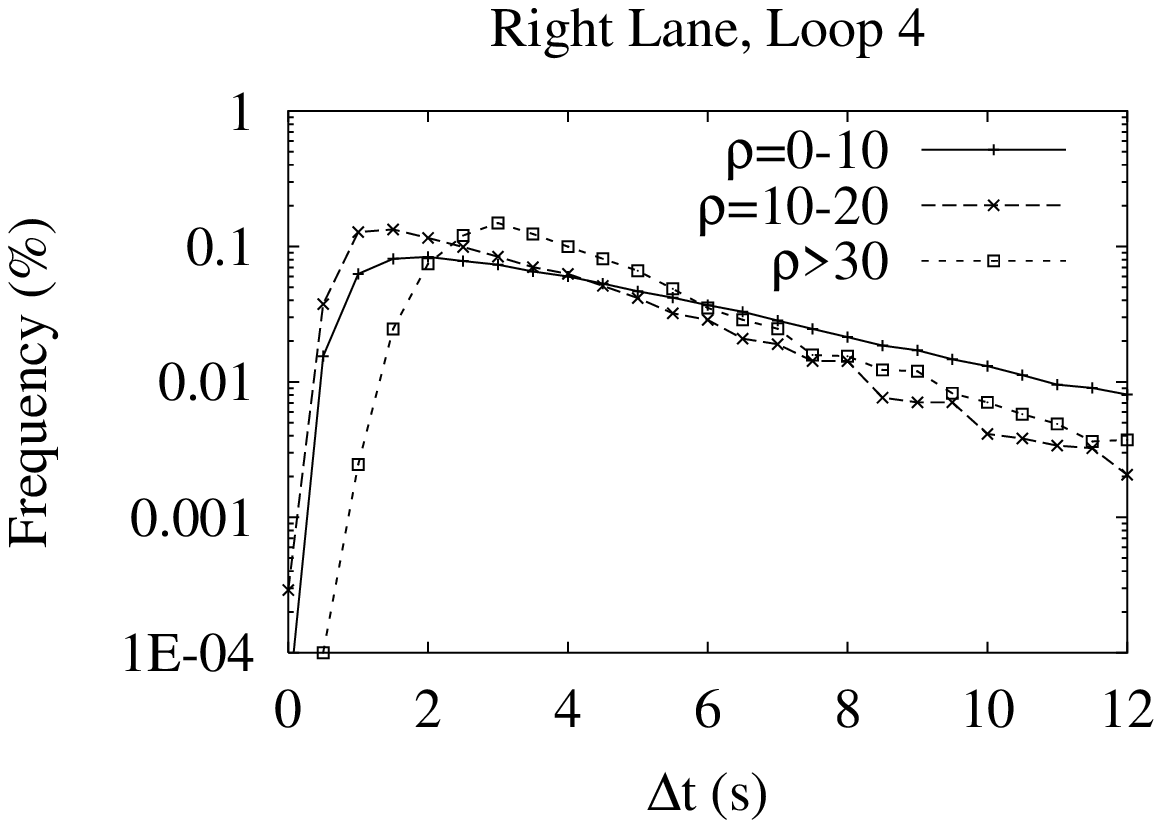}
\caption[]{\protect Half-logarithmically plotted time gap distributions 
for different density regimes, separately for the left and the right 
lane.\\[-14mm] \mbox{ }\label{fig:log}}
\end{center}
\end{figure}

In all cases we found practically continuous, unimodal time gap distributions.
The distribution for the right lane is usually broader than for the
left lane, and its maximum lies at higher time gaps $\Delta t$.
This is a consequence of the higher percentage of trucks.
The maximum for the left lane varies from 
$\Delta t_{\rm max}^{\rm free} \approx 0.9$\,s 
for $\rho_{50} = 0-20$\,vehicles/km 
up to $\Delta t_{\rm max}^{\rm congested} \approx 2.3$\,s for  
$\rho_{50} = 30-160$\,vehicles/km. Due to the existence of very large gaps,
the time gap distribution is broad at small densities. It is sharpest
immediately before the transition to congested traffic, and becomes
much broader afterwards. 
While, in the right lane, the time gap distributions decay exponentially 
for $\Delta t > 3$\,s, this seems to be different in the left lane 
(Fig.~\ref{fig:log}). Various suggestions for fitting the time gap
distributions can be found in Refs.~\cite{May90,Bovy98}.
\par
\begin{figure}
  \begin{center}
    \includegraphics[scale=0.43]{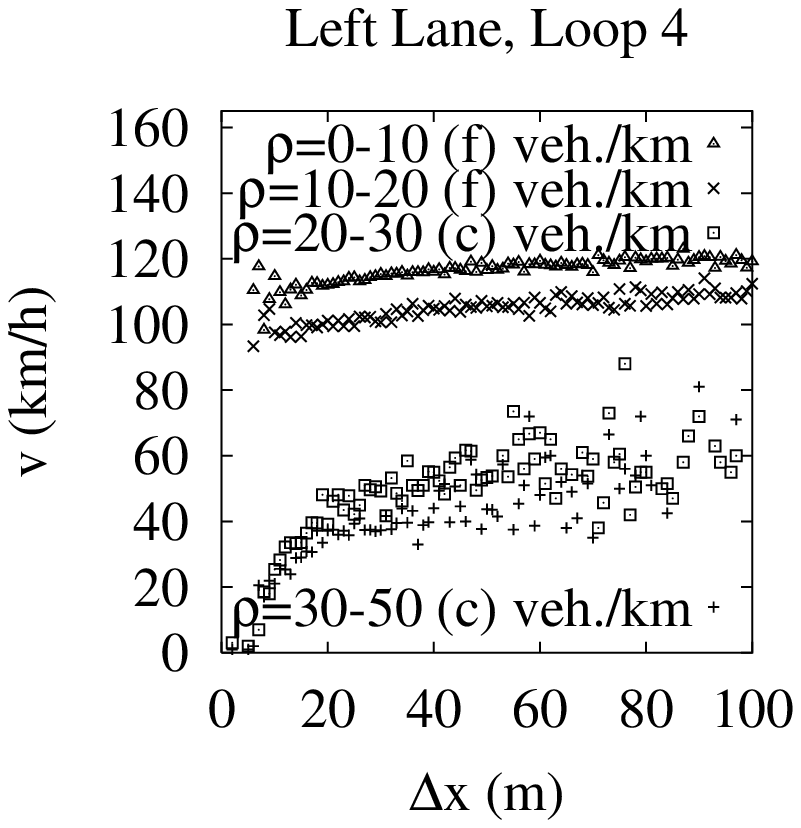}
    \includegraphics[scale=0.43]{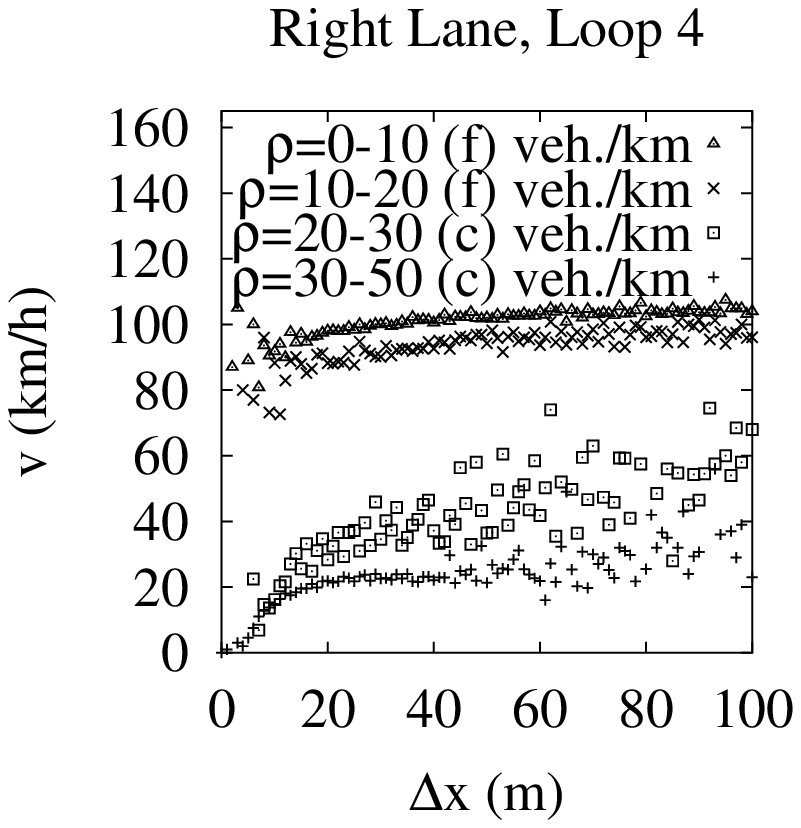}
    \includegraphics[scale=0.43]{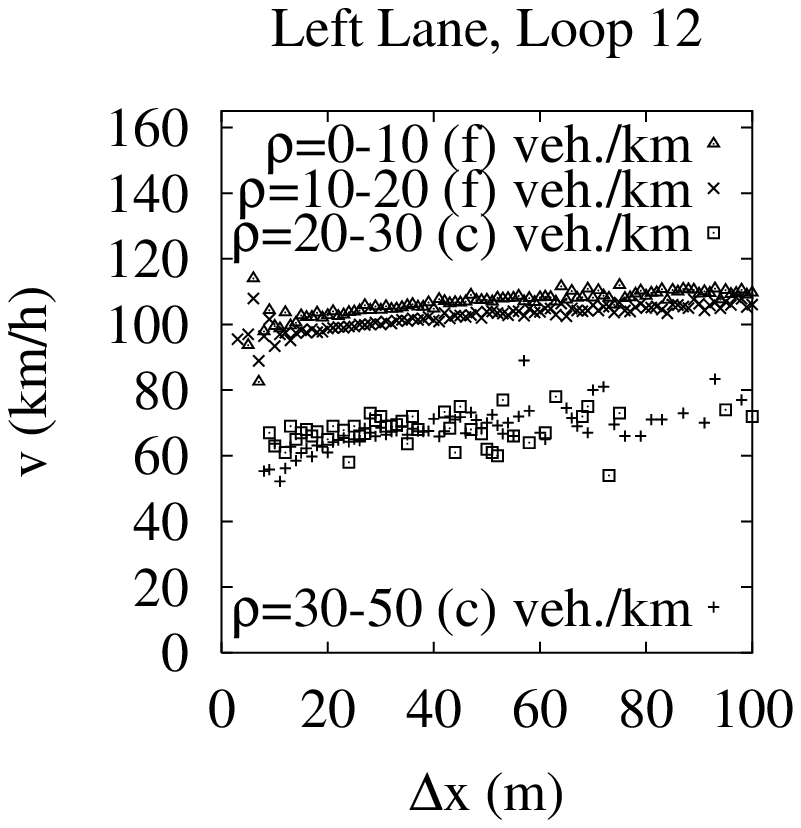}
    \caption[]{\protect Average vehicle velocities as a function 
of the distance gap for various density regimes, 
separately for the left and the right lane.\\[-12mm] 
\mbox{ }\label{fig:v_dx}}
\end{center}
\end{figure}
\begin{figure}[t]
  \begin{center}
     \includegraphics[scale=0.42]{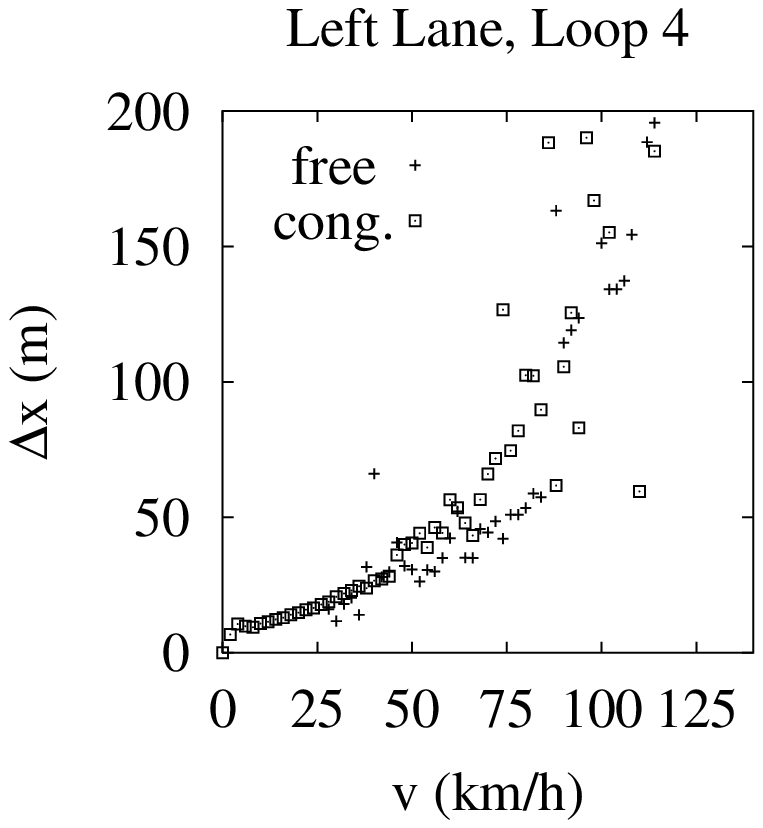}
     \includegraphics[scale=0.42]{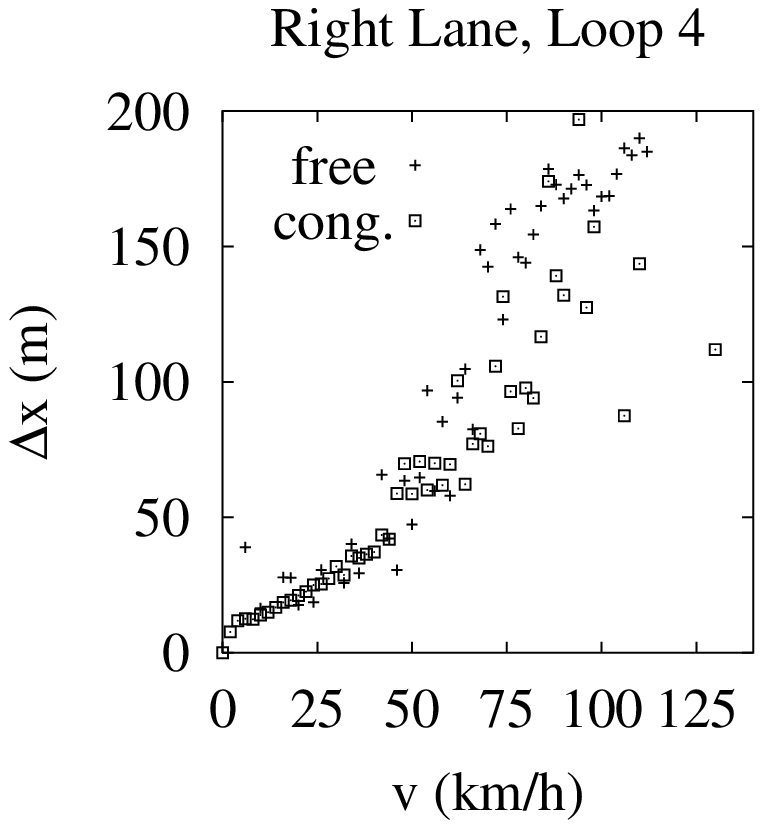}
     \includegraphics[scale=0.42]{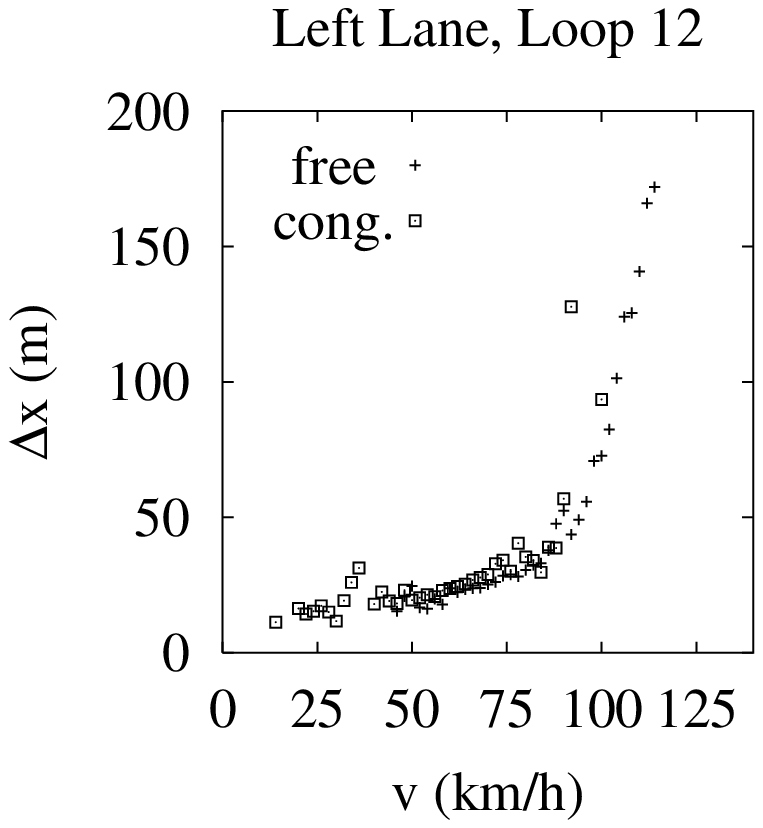}\\[-4mm]
\hspace*{8mm} \includegraphics[scale=0.5]{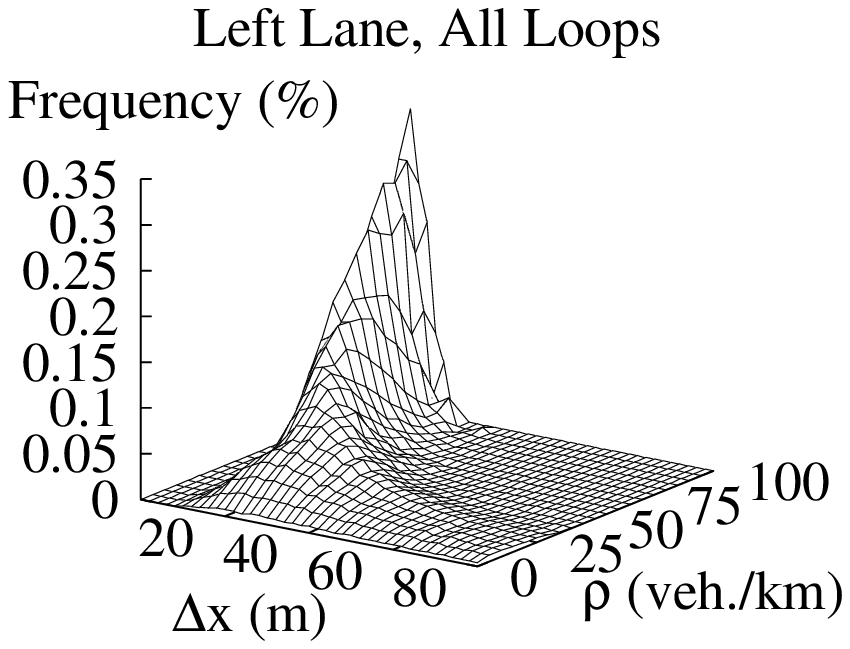}
    \includegraphics[scale=0.5]{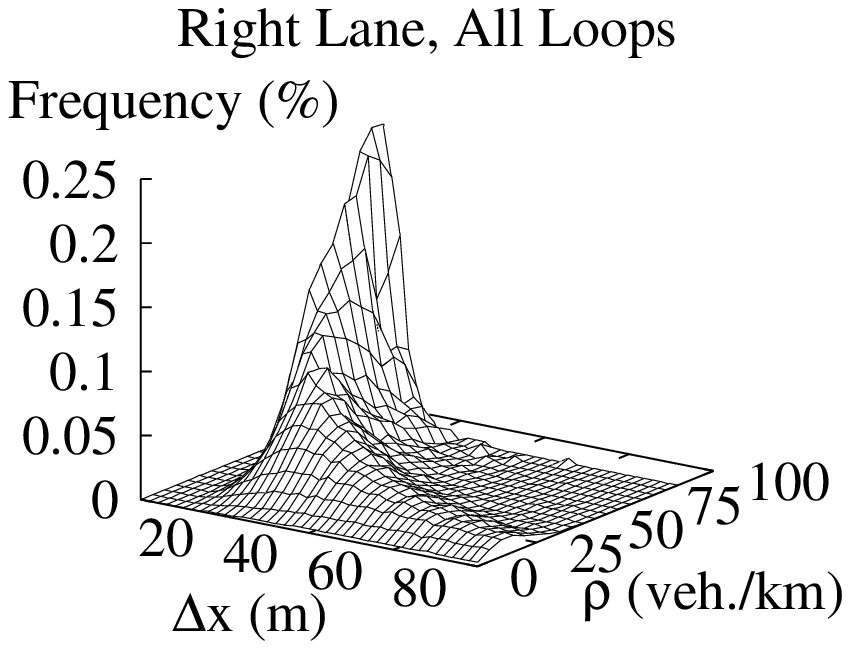}\\[-3mm]
    \caption[]{\protect Upper graphs: Average distance
gaps as a function of the velocity at different measurement sites.
Differences between free and congested traffic are found
upstream of a bottleneck (left, middle), but not 
away from it (right). Below: The broad
distance gap distributions change rather smoothly with
increasing density.\\[-12mm] \mbox{ }\label{fig:dx_v}}
\end{center}
\end{figure}

We have also plotted the average speeds $v$ of vehicles over their distance
headways $\Delta x$ to the respective car in front 
(Fig.~\ref{fig:v_dx}). The plots show the tendency of a 
density-dependent saturation of vehicle speeds with large
headways, which may be interpreted as frustration effect 
\cite{NeuSanSchaSchr99}. Like in Ref.~\cite{Bovy98},
we prefer to look at the data inversely
(Fig.~\ref{fig:dx_v}, upper graphs), since
the circumstance that we don't find typical headway-dependent velocities 
could just reflect large individual differences
in the ``optimal'' velocity-distance relations $v_i^{\rm d}(\Delta
x)$, or better: distance-velocity relations, 
which drivers prefer \cite{Ble98}. This would be consistent with the
broad distance gap distributions (Fig.~\ref{fig:dx_v}, lower
graphs). Note that some variation
of the distance gaps $\Delta x$ comes already from the fact that gaps with
regard to faster vehicles tend to be larger, because of the increasing
distance. Gaps with regard to slower vehicles tend to be larger as
well, since faster cars require some additional safety
distance to decelerate. (See Fig.~5 of 
Ref.~\cite{NeuSanSchaSchr99}.)\\{}\mbox{ }\\
{\it Acknowledgments:}
We would like to thank Henk Taale and the Dutch {\it Ministry of Transport, 
Public Works and Water Management} for supplying the freeway data,
and the BMBF for financial support (research
project SANDY, grant No.~13N7092).


\begin{thebibliography}{99}  

\bibitem
{May90}
May, A. D.: {\it Traffic Flow Fundamentals}
(Prentice Hall, Englewood Cliffs, NJ, 1990).

\bibitem
{Bovy98}
Bovy, P. H. L.: {\it Motorway Traffic Flow Analysis}
(Delft University Press, Delft, 1998).

\bibitem
{NeuSanSchaSchr99}
Neubert, L., {\it et al.}: Single-vehicle data of highway traffic---A
statistical analysis. Preprint cond-mat/9905216,
to appear in {\it Physical Review E} (1999).
\bibitem
{TreHel99}
Treiber, M. and Helbing, D.: Macroscopic simulation of widely
	scattered synchronized traffic states. {\it J. Phys. A:
	Math. Gen.} {\bf 32} (1999) L17-L23.

\bibitem
{Hel97} 
Helbing, D.: Traffic data and their implications for
consistent traffic flow modeling. In: 
{\it Transportation Systems}, edited by M. Papageorgiou and A. Pouliezos
(International Federation of Automatic Control,
Chania, Greece, 1997), Vol. II, pp. 809-814 (see cond-mat/9806221).

\bibitem
{HelBook}
Helbing, D.: {\it Verkehrsdynamik [Traffic Dynamics]} (Springer,
Berlin, 1997).

\bibitem
{Leu88}
Leutzbach, W.: {\it Introduction to the Theory of Traffic Flow}
(Springer, Berlin, 1988).

\bibitem
{Ble98}
Bleile, T.: {\it Modellierung des Fahrzeugfolgeverhaltens}
(PhD thesis, University of Stuttgart, 1999).
\end{thebibliography}
\end{document}